\documentclass[preprint2]{aastex}

\usepackage{rotating}
\usepackage{pdflscape}
\usepackage{mdwlist}

\shorttitle{New Northern \emph{NYMG} Members}
\shortauthors{Schlieder, L\'epine, and Simon}

\begin{document}

\title{Likely Members of the $\beta$ Pictoris and AB Doradus Moving Groups in the North}

\author{Joshua E. Schlieder\altaffilmark{1,2,3}, S\'{e}bastien L\'{e}pine\altaffilmark{4}, Michal Simon\altaffilmark{2,3}}
\altaffiltext{1}{Max-Planck-Institut f\"{u}r Astronomie, K\"{o}nigstuhl 17, 69117 Heidelberg, Germany, schlieder@mpia-hd.mpg.de}
\altaffiltext{2}{Department of Physics and Astronomy, Stony Brook University,
    Stony Brook, NY 11794, michal.simon@stonybrook.edu}
\altaffiltext{3}{Visiting Astronomer, NASA Infrared Telescope Facility (IRTF)}
\altaffiltext{4}{Department of Astrophysics,  American Museum of Natural History, Central Park West at 79th Street,
  New York, NY 10024, lepine@amnh.org}

\author{\emph{Accepted for publication in the Astronomical Journal}}

\begin{abstract}
 
We present first results from follow-up of targets in the northern hemisphere $\beta$ Pictoris and AB Doradus moving group candidate list of Schlieder, L\'epine, and Simon~(2012).  We obtained high-resolution, near-infrared spectra of 27 candidate members to measure their radial velocities and confirm consistent group kinematics.  We identify 15 candidates with consistent predicted and measured radial velocities, perform analyses of their 6-dimensional (\emph{UVWXYZ}) Galactic kinematics, and compare to known group member distributions.  Based on these analyses, we propose that 7 $\beta$ Pic and 8 AB Dor candidates are likely new group members.  Four of the likely new $\beta$ Pic stars are binaries; one a double lined spectroscopic system.  Three of the proposed AB Dor stars are binaries.  Counting all binary components, we propose 22 likely members of these young, moving groups.  The majority of the proposed members are M2 to M5 dwarfs, the earliest being of type K2.  We also present preliminary parameters for the two new spectroscopic binaries identified in the data, the proposed $\beta$ Pic member and a rejected $\beta$ Pic candidate.  Our candidate selection and follow-up has thus far identified more than 40 low-mass, likely members of these two moving groups.  These stars provide a new sample of nearby, young targets for studies of local star formation, disks and exoplanets via direct imaging, and astrophysics in the low-mass regime.                   

\end{abstract}

\section{Introduction}\label{intro}

Young stars are critical for understanding star formation and evolution and the circumstellar environment.  For example, the $\sim$10 Myr old A type dwarf $\beta$ Pictoris is host to the first directly imaged debris disk (Smith and Terrile~1984).  The first direct images of extrasolar planets were also achieved around nearby, young stars.  Images of the planetary system around the $\sim$30 Myr A type dwarf HR 8799 have provided empirical evidence to constrain planetary formation models and the evolution of the atmospheres in young, substellar objects (Marois et al.~2008; Marois et al.~2010; Barman et al.~2011, Skemer et al.~2012).  $\beta$ Pictoris also hosts an imaged massive planet whose orbit is consistent with an observed warp in the disk (Lagrange et al.~2010, 2012).  The previous stellar examples are members of the youngest populations of stars in the solar neighborhood; nearby, young moving groups (\emph{NYMGs}, Zuckerman and Song~2004, hereafter ZS04; Torres et al.~2008, hereafter T08).  Although moving group members have aged enough ($\sim$10-100 Myr) to be dispersed from their formation region, they retain small velocity dispersions and their common galactic motion translates to proper motions toward a convergent point in the plane of the sky (see Jones~1971; de Bruijne~1999; Mamajek~2005).

Because of their proximity (members $\lesssim$100 pc) and loosely bound nature, moving groups can be spread over thousands of square degrees on the sky.  Known groups include the AB Doradus and $\beta$ Pictoris moving groups ($\sim$50-150 and 10-20 Myrs old respectively), which are the only groups to have substantial numbers of known members in the north.  Tracing back the kinematics of the youngest \emph{NYMGs}, including $\beta$ Pic, shows that they may be related to a common star formation event in the Sco-Cen region (Mamajek, Lawson, and Feigelson~2000; Mamajek and Feigelson~2001, Fern\`andez et al.~2008, hereafter F08). In contrast, the older AB Dor group has kinematics and age similar to the Pleiades.  Luhman, Stauffer, and Mamajek (2005) suggest that AB Dor and the Pleiades are linked to the same star formation event.  

Member populations in these groups remain largely incomplete and searches to find the missing members are ongoing (e.g.~L\'epine and Simon~2009, hereafter LS09; Looper et al.~2010a, 2010b; Rice, Faherty, and Cruz~2010; Schlieder, L\'epine, and Simon~2010, hereafter SLS10; Kiss et al.~2011;  Riedel et al.~2011; Rodriguez et al.~2011; Shkolnik et al.~2011; Zuckerman et al.~2011; Malo et al.~2012; Schlieder, L\'epine, and Simon~2012, hereafter SLS12; Shkolnik et al.~2012).  Low-mass members (M$\lesssim$1.0 M$_{\odot}$) remain the most under sampled population because: (1) the intrinsic luminosity of low-mass stars places them beyond the flux limits of the X-ray and astrometric catalogs that have served as the primary resources for identifying new group members (see ZS04; T08 and references therein), and (2) Youth diagnostics for the lowest-mass stars lack reliability, this makes the missing stars more difficult to identify from the field (see Schlieder et al.~2012, hereafter S12).  

The low-mass populations of moving groups are key for studies of recent local star formation, disks and exoplanets, and the astrophysics of young, low-mass stars.  With a more complete moving group census the mass functions of individual groups could be determined, allowing direct comparison to young open clusters and the field (e.g. Kroupa et al.~2002, Chabrier et al.~2005, Bochanski et al.~2010).  Studies of this kind could provide insight into the universality of the initial mass function and reveal information about the formation environment of \emph{NYMGs}.   Since moving groups may represent the endpoint of clustered star formation, they are crucial not only for understanding the recent star formation history near the Sun, but the latter stages of cluster evolution and dispersion in general.  A complete sample of nearby, young stars would also allow for detailed multiplicity studies.   Multiplicity fractions, mass-ratios, and semi-major axis distributions can also be linked to formation environment and dynamical evolution (see Chauvin et al.~2010; Bergfors et al.~2010; Janson et al.~2012, hereafter J12, and references therein).  In addition, identification of low-mass moving group members will allow for the detailed investigation of signatures of youth in low-mass stars.  These include chromospheric and coronal activity indicators, lithium abundances, gravity sensitive spectral features, and rapid rotation.  A more complete sample would be particularly useful in understanding the evolution and reliability of these youth signatures at different ages across the transition to the fully convective regime ($\sim$M4 spectral type, see West et al.~2011, hereafter W11; Reiners and Mohanty~2012, hereafter RM12; Reiners, Nandan, and Goldman~2012; S12).                   
  
Observations of the few known, low-mass \emph{NYMG} members have provided a wealth of information regarding circumstellar disks and substellar companions.  The most classic example may be AU Mic, a $\sim$10 Myr old M dwarf in the $\beta$ Pic group, which hosts an edge on debris disk that has acted as a benchmark for disk evolution in the low-mass regime (Kalas et al.~2004, Liu~2004).  Known low-mass members host to substellar companions include 2MASS 1207 in the TW Hydrae association, an $\sim$8 Myr old brown dwarf with a planetary mass companion (Chauvin et al.~2004), and CD-35 2722 and 1RXS J2351+3127, AB Dor M dwarfs with brown dwarf companions (Wahhaj et al.~2011; Bowler et al.~2012, hereafter B12).  Further imaging surveys of low-mass moving group stars will provide statistics on the fraction of M dwarfs with massive planet and brown dwarf companions (see B12; Delorme et al.~2012).  These observations can infer key constraints on multiple system and planet formation in the low-mass regime.  

A new technique to identify low-mass moving group members was described in LS09 and SLS10.  Low-mass candidate group members are selected on the basis of proper motion, photometry, and activity.  In SLS12 we described a comprehensive list of young $\beta$ Pictoris and AB Dor candidates in the northern hemisphere selected using this technique.  The use of new proper motion catalogs and UV flux as an activity indicator allowed access to large numbers of previously inaccessible, young candidates.  These stars require detailed follow-up to verify group membership.  In this paper we present our first results from a multi-epoch, multi-telescope follow-up campaign of this candidate list.  In \S 2 we describe our observations and present the results.   An analysis of the 6-dimensional (6D, \emph{UVWXYZ}) kinematics of proposed likely members is presented in \S 3.  Likely new members (\emph{LNMs}) are treated individually in \S 4 and two new spectroscopic binaries from the sample are discussed in more detail in \S 5.  We present a discussion of proposed member distances and ages in \S 6 and a summary follows in \S 7.

\section{Observations and Results}

We have observed 27 northern, probable young candidates (\emph{PYCs}) using the \emph{Cryogenic Echelle Spectograph} (CSHELL, Greene et al.~1993) on the 3.0-m NASA \emph{Infrared Telescope Facility} (IRTF) and \emph{Phoenix} on the 8.1-m \emph{Gemini-South} telescope to measure their radial velocities (RVs) (see Table \ref{PYC_NORTH}).  All CSHELL  measurements were made using the same instrument settings, and extracted using the same techniques, described in SLS10.  \emph{Phoenix} is a high resolution near-IR echelle spectrometer (Hinkle et al.~2003).  Observations are acquired over a very small range of wavelengths (approximately 0.5\% of the central wavelength) since no cross-disperser is available.  Single order spectra are obtained by using blocking filters.  We used the instrument centered at 1.5548 $\mu$m with the H6420 blocking filter, which transmits light in the range 1.547 $\le$ $\lambda$ $\le$ 1.568.  To achieve a resolution of  R$\sim$50,000, we used the 0\farcs34 slit.  Calibration observations included standard flats, darks, and wavelength calibration.  The \emph{Phoenix} data were extracted, after cosmic ray removal, using the same software as the CSHELL data.   We measured candidate RVs, rotational velocities ($v$sin$i$), and spectral types (SpTy) with the same software and cross-correlation techniques described in SLS10.           

Table~\ref{PYC_NORTH} lists the 12 $\beta$ Pic and 15 AB Dor \emph{PYCs} targeted for follow-up.  Columns 1 and 2 are the candidate's northern list ID and alternate name.  Columns 3 and 4 list the \emph{IRCS} (J2000) coordinates and columns 5 and 6 are the \emph{V} and \emph{K$_s$} magnitudes.  \emph{PYC} kinematic distances ($d_{kin}$), model distances ($d_{mod}$), predicted and measured RVs (RV$_p$, RV$_m$), and $v$sin$i$ are listed in columns 7-11.  Columns 12 and 13 list the SpTy of the candidate and if the star's RV$_m$ is consistent with its RV$_p$ within the tolerance described in SLS10; indicating likely group membership.

Seven of the $\beta$ Pic candidates observed have measured RVs consistent with group membership.  Four of these stars are multiples:  PYC J00325+0729AB is a $<$1$^{\prime\prime}$ visual binary, PYC J07295+3556AB is an $\sim$0.2$^{\prime\prime}$ binary discovered by lucky imaging,  PYC J10596+2526E is the secondary component in a $\sim$5$^{\prime\prime}$ visual binary, and multiple RV measurements show that PYC J09362+3731AB is an equal mass, double lined, spectroscopic binary (SB2).  One rejected $\beta$ Pic candidate, PYC J22187+3321AB, is also an SB2.  These newly identified SB2s are detailed in \S\ref{SB2}.  Eight of the AB Dor candidates also have consistent RVs, including 3 visual binaries with separations ranging from $\sim$1$^{\prime\prime}$ - $\sim$4$^{\prime\prime}$.  Most of the \emph{PYCs} observed are M dwarfs, a result of more sensitive proper motion catalogs searched to generate the northern list (see SLS12).

\section{Kinematic Analysis}\label{KINEMATICS}

Members of young moving groups are defined by their common galactic kinematics in 6D (\emph{UVWXYZ}) coordinates.  In this system, \emph{U}, \emph{V}, and \emph{W} are Galactic space velocities relative to the Sun (see Johnson and Soderblom 1987, hereafter JS87).  We define them to be positive in the direction of the Galactic center, Solar motion around the Galaxy, and Galactic north pole respectively.  \emph{X}, \emph{Y}, and \emph{Z} are Galactic distances defined positive in the same directions as (\emph{UVW}).  In this section, we compare the Galactic kinematic distributions of known $\beta$ Pic and AB Dor members to the young candidates having consistent measured and predicted RVs (RV match candidates).  We use the methods outlined by JS87 to calculate the Galactic kinematics and their uncertainties.  For the known members, (\emph{UVWXYZ}) parameters were calculated from data available on \emph{SIMBAD} and in T08, Torres et al.~(2006), and F08.  For RV match candidates, data was compiled from the SLS12 list, \emph{Hipparcos} and \emph{Tycho2} when available, and our RV measurements.  We list the calculated kinematics for the candidates in Table~\ref{6D_DATA}.

\subsection{Model Derived Distances}

As part of the selection process in SLS12, we verified that the kinematic distances and photometry of our young candidates are consistent with the cluster sequences of known $\beta$ Pic and AB Dor members when placed in  color-magnitude-diagrams.   As another consistency check, we estimate model derived photometric distances.  We first converted RV match candidate (\emph{V-K$_{s}$}) colors to effective temperatures using the conversion scale of Kenyon and Hartmann~(1995).  We then used these temperatures to extract model derived \emph{K} magnitudes from Baraffe et al.~(1998) pre-main-sequence evolution models and estimated their distances from the distance modulus\footnote{Kenyon and Hartmann~(1995) and Baraffe et al. use Bessel \& Brett~(1988) and CIT near-IR magnitudes respectively. We used the \emph{2MASS} (Skrutskie et al.~2006) magnitude conversions given at http://www.astro.caltech.edu/~jmc/2mass/v3/transformations/ to place all photometry on the \emph{2MASS} scale.}.  

For the $\beta$ Pic and AB Dor candidates respectively, we estimated their distances for ages of 10, 20, and 30 Myr  and 50, 70, and 100 Myr to reflect group age estimates from the literature (see F08).  We applied a correction factor of (1+ $\zeta$)$^{1/2}$ to the model derived distances of binary candidates with unresolved photometry, where $\zeta$ is the \emph{K$_{s}$} magnitude ratio of the components.  In the cases where the SpTy of the secondary is known but no photometry is available, we estimate the mag. ratio of the components using the models.  In cases where there is neither a SpTy or photometric data for the secondary, we assume the system to have equal flux components.  

We found that 20 Myr model distances best match the $d_{kin}$ of the $\beta$ Pic candidates.  We estimate a systematic uncertainty of $\sim$50 \% from the range of model distances derived at different ages.  50 Myr models provide the best match to the AB Dor candidate $d_{kin}$ with an uncertainty of $\sim$20 \%.  Within uncertainties, all candidate kinematic distances are consistent with model distances at the nominal ages of the groups.  We thus adopt candidate $d_{kin}$ as the distance used in the kinematic analysis.  The $d_{mod}$ for the RV match candidates of $\beta$ Pic and AB Dor, at 20 and 50 Myr respectively, are listed in column 8 in Table~\ref{PYC_NORTH}.   

\subsection{$\beta$ Pic Group Kinematics}

\begin{figure*}[!h]
\begin{center}
\includegraphics[width=6.6truein]{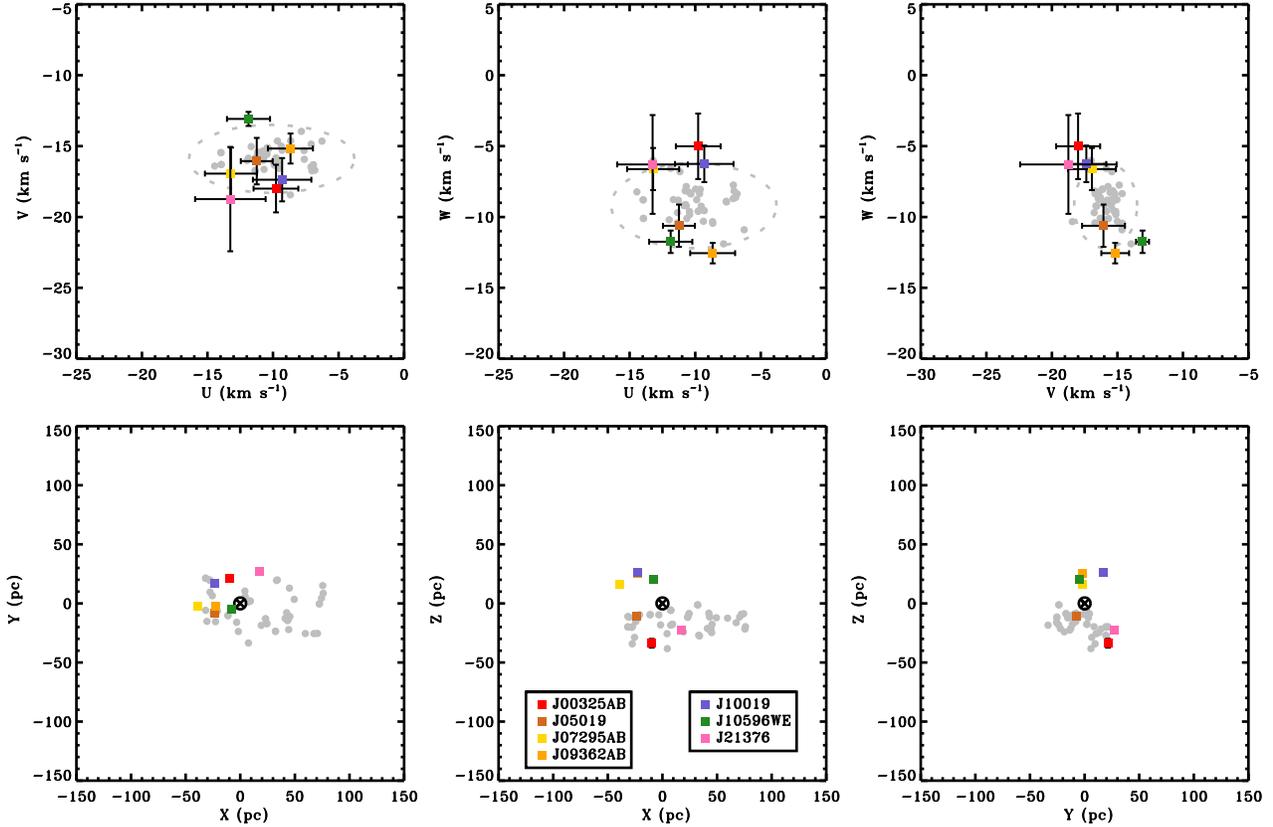}
\caption[]{(\emph{UVWXYZ}) projections of known and candidate $\beta$ Pictoris moving group members.  Known members, shown as gray, filled circles, are taken from the membership list proposed by T08.  The top 3 panels are projections of (\emph{UVW}) space velocities and 3$\sigma$ error ellipses are shown as dashed gray lines.  The candidate members with consistent measured and predicted RVs are shown as colored, solid squares with individual error bars.  The color designations are listed in the legend.  Each of the candidate systems are consistent with the (\emph{UVW}) distributions of the known group members.  In (\emph{XYZ}) space, showed in the bottom panels, several proposed likely members extend to +$Z$ distances.  These stars occupy a part of the sky where $\beta$ Pic members were not previously known (see Fig.~\ref{SKY}).  The black cross at (\emph{XYZ}) = (0, 0, 0) represents the Sun.\\ \\(A color version of this figure is available in the online journal.)}
\label{BP_6D}
\end{center}
\end{figure*}

Figure~\ref{BP_6D} shows the 6D kinematic distributions of known and candidate $\beta$ Pic members.  The top 3 panels are projections in (\emph{UVW}) velocity space, the bottom 3 are equivalent projections in (\emph{XYZ}).  Known members of the group taken from T08\footnote{Some known members of $\beta$ Pic and AB Dor did not have complete astrometric data in the literature and are not shown in the (\emph{UVW}) projections.}, are shown as gray, filled circles, and have mean (\emph{UVW})$_{BP}$ = (-10.1$\pm$2.1, -15.9$\pm$0.8, -9.2$\pm$1.0) km s$^{-1}$ and mean (\emph{XYZ})$_{BP}$ = (20, -5, -15) pc (T08).  As evidenced by the figure, the range of distances is large, nearly 100 pc in \emph{X}.  The  3$\sigma$ (\emph{UVW}) error ellipses are shown as dotted lines.  We define these regions to be the acceptable range of (\emph{UVW}) velocities for proposed new group members.   

The candidates are shown as colored, filled, squares.  Their designations can be found in the figure legend.  In the (\emph{UVW}) projections in the top panels, within uncertainties, all of the candidates exhibit Galactic kinematics that are consistent with the known member distributions.  In the bottom panels of Figure~\ref{BP_6D}, four candidates have positive \emph{Z}~distances.  This is inconsistent with the \emph{Z} distribution of previously known members.  However, these candidates occupy a region of the sky where $\beta$ Pic members were not previously known (see Fig.~\ref{SKY}).  Since our search is one of the first to concentrate on northern targets, it is possible that these outliers are a natural extension of the group into a poorly searched coordinate space that corresponds to +\emph{Z}~distances.  For these reasons, we do not reject these candidates as likely members based on this discrepancy.  Future follow-up of the SLS12 candidate list and other all sky searches for new $\beta$ Pic members may contribute more examples to this subpopulation.  Our kinematic analysis of $\beta$ Pic RV match candidates reveals that all 7 systems are consistent with the expected (\emph{UVW}) distribution, we thus retain each candidate as a likely new member of the $\beta$ Pic group and describe them individually in \S\ref{NORTH_LNM}.

\subsection{AB Dor Group Kinematics}

\begin{figure*}[!h]
\begin{center}
\includegraphics[width=6.6truein]{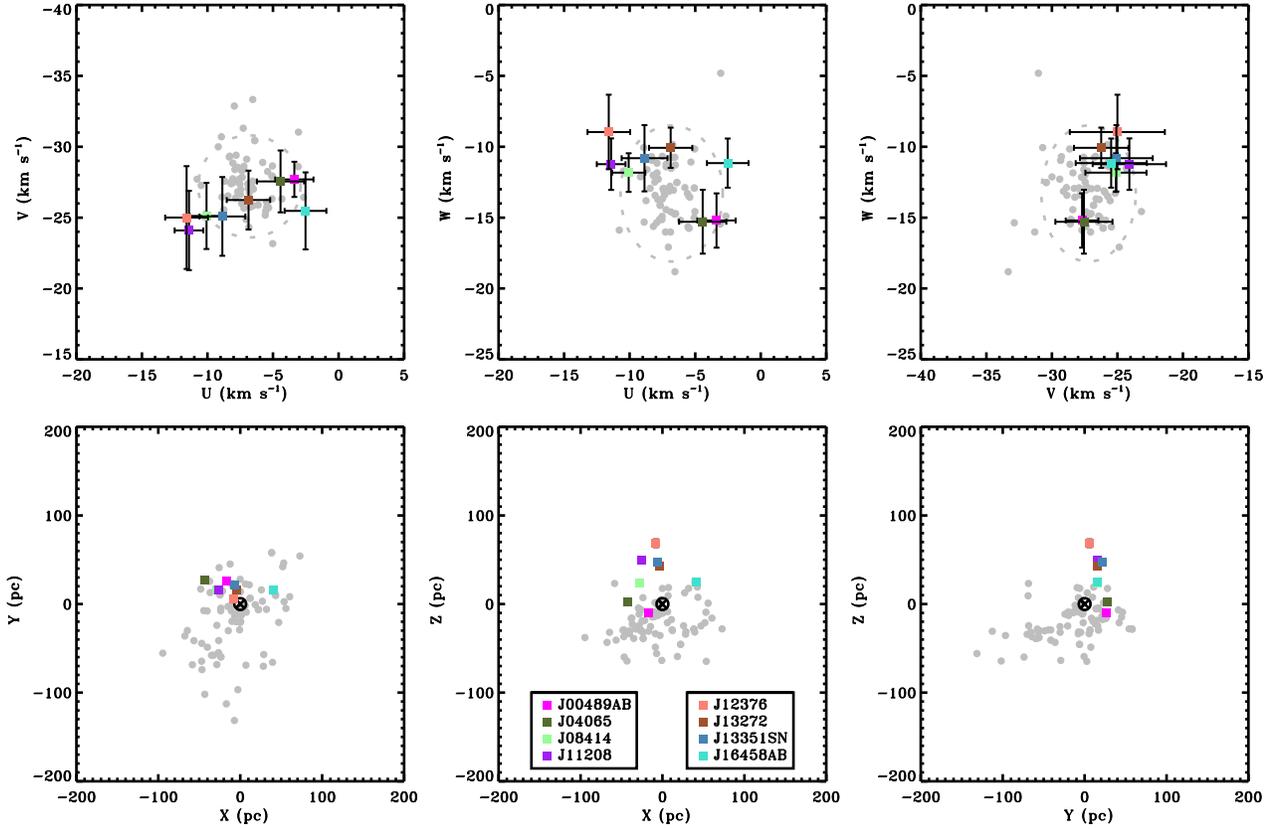}
\caption[]{(\emph{UVWXYZ}) projections of known and candidate AB Doradus moving group members.  Known members are from T08.  Panel and symbol designations are identical to Figure~\ref{BP_6D}.  All 8 of the RV match candidates are consistent with the expected (\emph{UVW}) distributions of the group.  Our focus on the northern hemisphere has identified likely members at larger $Z$ distances than previous searches.\\ \\(A color version of this figure is available in the online journal.)}
\label{ABD_6D}
\end{center}
\end{figure*}

\begin{figure*}[!h]
\begin{center}
\includegraphics[width=3.0truein]{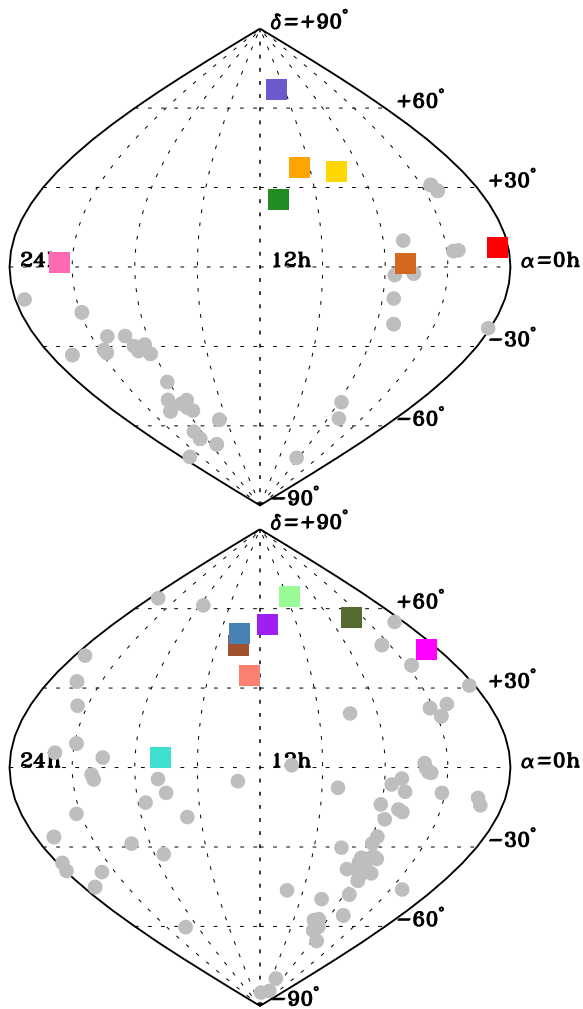}
\caption[]{Sinusoidal sky projections showing positions of known and proposed new $\beta$ Pictoris (top) and AB Doradus (bottom) moving group members (known members from T08).  Symbol and color designations are identical to Figures~\ref{BP_6D} and~\ref{ABD_6D}.  Proposed new $\beta$ Pic members with coordinates 9h$<$$\alpha$$<$ 2h and $\delta$$> $20$^{\circ}$ occupy a region of the sky  where members of the group were not previously known.  These stars are also the first proposed members to lie at $Z$$>$0 pc (see Fig.~\ref{BP_6D}).  Some proposed AB Dor members around $\alpha$=12h and with $\delta$$>$30$^{\circ}$ also have positions and distances that correspond to new regions of +$Z$ space for this group (see Fig.~\ref{ABD_6D}).  Although we have observed several young $\beta$ Pic candidates in the north between 12h and 24h, we find few likely members.  AB Dor members occupy the entire sky.\\ \\(A color version of this figure is available in the online journal.)}
\label{SKY}
\end{center}
\end{figure*}

Figure~\ref{ABD_6D} shows the 6D kinematic distributions of known and candidate AB Doradus members.  The panel and symbol designations are the same as for Figure~\ref{BP_6D}. The mean velocities and distances of known AB Dor members from T08 are (\emph{UVW})$_{ABD}$ = (-6.8$\pm$1.3, -27.2$\pm$1.2, -13.3$\pm$1.6) km s$^{-1}$ and (\emph{XYZ})$_{ABD}$ = (-6, -14, -20).  The distance range between given members can be very large, over 150 pc (T08).  In the top panels, all candidates are again consistent with the acceptable regions of (\emph{UVW}) space for AB Dor membership within uncertainties. 

In the bottom panels, the known AB Dor members are spread over a much larger volume of space than those of the younger $\beta$ Pic group.  This is likely due to longer dynamical mixing and more interactions over the lifetime of the group.  The candidates are generally consistent with the known member (\emph{XYZ}) distributions, although there are again several that lie at +\emph{Z} distances greater than previously known members.  Similar to the results for the $\beta$ Pic candidates, we have identified candidates that occupy an expanded region of (\emph{XYZ}) space by dedicating our search to northern declinations (see Fig.~\ref{SKY}).  These results are in support of all 8 AB Dor RV match candidate systems being likely new members of the group.  They are treated individually in \S\ref{NORTH_LNM}.

\section{Northern Likely New Members}\label{NORTH_LNM}

\subsection{$\beta$ Pictoris Moving Group}

\noindent{\emph{PYC~J00325+0729AB = LP~525-39AB}:  LP~525-39AB is a pair of M4 dwarfs at a predicted distance of 41.1 pc.  The system was identified as a 0.79$^{\prime\prime}$ binary in a survey of nearby dwarfs to search for substellar companions (McCarthy, Zuckerman, and Becklin~2001, hereafter M01).  We identify strong X-ray, NUV, and FUV emission; consistent with young, active chromospheres.  The RV of the secondary was previously measured to be -4.6 km s$^{-1}$, this is consistent with our RV measurement of both components (Gizis, Reid, and Hawley~2002).  The large $v$sin$i$ is also consistent with youth.  Several photometric distance estimates can be found in the literature and all estimate the distance to be $\sim$12 pc (M01; Weis et al.~1986; L\'epine and Gaidos~2011, hereafter LG11).  These distances are inconsistent with our kinematic and model derived estimates, however they are based on unresolved photometry.} 
\newline

\noindent{\emph{PYC~J05019+0108}:  We identify PYC~J05019 +0108 as an active M4 dwarf a predicted distance of 27.0 pc.  The star exhibits strong X-ray emission consistent with other young M dwarfs, but is not listed in the \emph{GALEX} All-Sky Imaging Survey (Morrissey et al.~2005).  S12 measured strong H$\alpha$ emission and a weak Na 8200~\AA~doublet, attributed to low surface-gravity.  The equivalent width of the sodium doublet is consistent with an age of $<$100 Myr.  We also measure a large $v$sin$i$ consistent with youth.  LG11 estimate a photometric distance of 10.4$\pm$3.1 pc.  This estimate is inconsistent with our kinematic and model derived distances and parallax measurements are needed to verify $\beta$ Pic membership.}      
\newline

\noindent{\emph{PYC J07295+3556AB}: PYC J07295+3556AB is an M1/M3 binary with a separation of 0.198$^{\prime\prime}$ at a predicted distance of 42.2 pc.  The binary nature of the system was discovered in a lucky imaging survey of active M dwarfs (J12).  We estimate a combined spectral type of M4 from 4 epochs of H-band spectra of the system.  The pair exhibits strong X-ray, NUV, and FUV emission.  We also measure a moderately high rotational velocity, although this could be due to blending in the combined spectrum.  It was previously identified as an M1 dwarf at a spectroscopic distance of 37 pc (Riaz, Gizis, and Harvin~2006, hereafter R06).  This distance estimate is consistent with our kinematic and model derived estimates.  J12 estimate the separation of the components to be 8.5$\pm$3.1 AU using the Riaz et al.~distance.  This corresponds to an $\approx$38 yr orbital period.  In fact, in 5 imaging epochs spanning 1.5 years, J12 measure a 4$^{\circ}$ change in position angle.  Thus, the full orbit and dynamical masses for this pair can be determined with continued monitoring.  These likely $\beta$ Pic members may eventually prove to be a valuable test to evolution models in the low-mass regime.}
\newline

\noindent{\emph{PYC J09362+3731AB = HIP 47133AB}:  HIP 47133AB was first proposed as a likely new member of the $\beta$ Pic moving group in SLS12 on the basis of consistent proper motion, strong activity, and consistent trigonometric distance.  We have now obtained three RV measurements and find it to be an M2/M2 spectroscopic binary having a center-of-mass velocity consistent with $\beta$ Pic membership.  This system may represent a rare opportunity to measure dynamical masses of a young, low-mass binary at a measured distance for comparison to evolution models.  The observations, system parameters from available data, and their implications are discussed in detail in the following section.} 
\newline

\noindent{\emph{PYC J10019+6651}:  We identify PYC J10019 +6651 as an M3 dwarf at a predicted distance of 38.7 pc.  In SLS12, this star is listed as a candidate of both the $\beta$ Pic and AB Dor moving groups.  The measured RV  is consistent with the predicted RV for $\beta$ Pic membership over multiple epochs.  The dwarf exhibits strong X-ray, NUV, and FUV emission consistent with other young, late-type stars.  LG11 estimate a photometric distance of 24.6$\pm$7.4 pc.  This is broadly consistent with our kinematic and model estimates.  If confirmed as an unambiguous member, this star will be the northernmost member of the $\beta$ Pic moving group.}    
\newline

\noindent{\emph{PYC J10596+2526E = BD+26 216B}:  BD+26 216B is the K5 secondary of HD 95174, a K2 dwarf.  The primary was cut during the candidate search of SLS12 because it has (\emph{V-K$_s$})$<$3.2.  The components are separated by $\sim$5$^{\prime\prime}$.  The system has a spectroscopic distance estimate of 23.2 pc, which is consistent with our $d_{kin}$ and $d_{mod}$ (Scholz, Meusinger, and Jahreiss~2005).  The system does not have a \emph{ROSAT} counterpart and was selected as a \emph{PYC} on the basis of strong UV emission.  The components are unresolved in \emph{GALEX} images but the source does exhibit some elongation.  Since the NUV/\emph{K$_s$} and FUV/\emph{K$_s$} flux ratios listed in SLS12 for BD+26 2161B are based on the integrated UV flux of the system, the individual emission is not as strong as first thought.  However, the K-type stars still exhibit moderate UV emission and the galactic kinematics of the system do match those of the $\beta$ Pic group.  Estimates of youth from spectroscopy should confirm the membership of this system.  At the age of $\beta$ Pic, the early-K primary should still have photospheric lithium.  We consider both the primary and secondary  likely new members}        
\newline

\noindent{\emph{PYC J21376+0137 =  EUVE J2137+01.6}:  We identify EUVE J2137 +01.6 as an active M5 dwarf at a predicted distance of 39.2 pc.  The star exhibits strong X-ray, NUV, and FUV emission; driven by its large rotational velocity of $\sim$50 km s$^{-1}$.  S12 also measure strong H$\alpha$ emission and a weak, gravity sensitive, sodium doublet in the red optical.  There are several RV measurements of this star available in the literature and we have measured the RV at four different epochs in September and October 2009.  The range of measured RVs is large; from -8.8 to -15.1 km s$^{-1}$ (see Fleming~1998; Mochnacki et al.~2002, hereafter M02; this work).  Treating the star as an SB2 produces unphysical results in the cross-correlation, so it is likely not an unresolved binary.  The variation in measured RV is probably a result of the rapid rotation which leads to large uncertainties in the measurements.  EUVE J2137+01.6 was also previously detected as a UV source, stellar radio source, and found to exhibit a strong magnetic field from spectropolarimetric observations (Christian et al.~1999; Helfand et al.~1999; Phan-Bao et al.~2009).  Photometric distance estimates in the literature are discrepant with our kinematic and model estimates; predicting the star to lie at $\sim$10 pc.  Parallax measurements are needed to verify that this highly active and intriguing M dwarf is a member of the $\beta$ Pic moving group.}

\subsection{AB Doradus Moving Group}

\noindent{\emph{PYC J00489+4435EW = LP 193-584AB}:  LP 193-584AB is a pair of mid-M dwarfs separated by 1.09$^{\prime\prime}$ (M01).  The pair exhibits strong X-ray emission but was not observed by \emph{GALEX}.  We measure a spectral type of M4 from the combined light spectrum of both components.   We estimate a distance of 32.6 pc, which is very comparable to the spectroscopic distance estimate of 33 pc from R06.  M02 measure RV=-18.4$\pm$1.8 km s$^{-1}$ and $v$sin$i$=36 km s$^{-1}$, their RV is consistent with our measurement but their rotational velocity is larger.  It may be affected by blending of the components.}
\newline

\noindent{\emph{PYC J04065+5628 = TYC 3726 1068 1}: We identify TYC 3726 1068 1 as an X-ray, NUV, and FUV active K7 at about 50 pc. The only mention of this star in the literature is a previous designation as a \emph{ROSAT} (Voges et al.~1999, 2000) X-ray source (Haakonsen and Rutledge~2009).  Future spectroscopic observations will provide more information and confirm the star's suspected group membership.}  
\newline

\noindent{\emph{PYC J08414+6425 = NLTT 19987}:  NLTT 19987 is an active M2 dwarf at a predicted distance of 39.9 pc.  There is little information in the literature about this proposed member, aside from the strong X-ray, NUV, and FUV emission quoted in SLS12 and a photometric distance of 38.2$\pm$11.5 pc from LG11. This distance is consistent with our kinematic and model estimates. If confirmed a true member of the group it will be the northernmost member.}  
\newline

\noindent{\emph{PYC J11208+5410 = TYC 3825 716 1}:  We identify TYC 3825 716 1 as an active K7 dwarf at 57.9 pc.  The star exhibits strong X-ray, NUV, and FUV emission consistent with youth.  LG11 estimate a photometric distance of 49.8$\pm$14.9 pc; consistent with our estimates.}  
\newline

\noindent{\emph{PYC J12376+3450 = TYC 2533 1877 1}:  We measure that TYC 2533 1877 1 is a K7 and estimate a distance of 69.4 pc.  The star was previously identified as a counterpart to a \emph{ROSAT} source and we find that it also exhibits strong UV emission (Zickgraff et al.~2003; SLS12).  The star has a spectroscopic distance estimate of 68 pc which is consistent with our predicted distance (R06).}
\newline

\noindent{\emph{PYC J13272+4558 = TYC 3460 416 1}:  We identify TYC 3460 416 1 as an active K7 dwarf at an estimated distance of 46.0 pc.  The star is a counterpart to a \emph{ROSAT} source and is a photometric variable with a period of 2.183 d (Norton et al.~2007).  There is little information about this star in the literature and further follow-up will confirm its proposed membership in the AB Dor group.}  
\newline

\noindent{\emph{PYC J13351+ 5039S and N}:  PYC J13351+5030 was found to be a $\sim$4$^{\prime\prime}$ visual binary in CSHELL acquisition images.  We were able to measure the RV of both components and they are consistent within uncertainties.  The large $v$sin$i$ of the components and activity from SLS12 are consistent with a young age.  The system lies at a kinematic distance of 52.8 pc; our model derived distance, while based on poor quality \emph{2MASS} photometry, is consistent with this estimate.}
\newline

\noindent{\emph{PYC J16458+0343N and S}:  PYC J16458+0343 was identified as a $\sim$1$^{\prime\prime}$, M3/M4 visual binary during CSHELL observations.  We were only able to measure the RV of the brightest component, but we consider them a visual binary .  The system was previously identified as a counterpart to a \emph{ROSAT} X-ray source, SLS12 also finds it is also bright in the UV.  LG11 estimate a photometric distance of 30.6$\pm$9.2 pc.  This estimate is based on unresolved photometry but is still broadly consistent with our kinematic and model distance estimates.}

\section{Spectroscopic Binaries}\label{SB2}

Observations of spectroscopic binaries allow for the measurement of the radial component of binary motion through the Doppler shift of spectral features.  Many parameters of the binary orbit can be determined from this measurement, including but not limited to, component masses, the period, and the center-of-mass velocity of the system.  In a double lined system, where spectral features of both components allow for individual RV measurements, the mass ratio of the stars, $q$ = M$_2$/M$_1$,  and center-of-mass velocity, $\gamma$, can be calculated dynamically from just 2 observations, $i$ and $j$ (Wilson 1941).  The $q$ and $\gamma$ are given by

\begin{equation}
q = -{{V_{1,i} - V_{1,j}}\over{V_{2,i} - V_{2,j}}}
\end{equation}

\noindent{and} 

\begin{equation}
\gamma={{V_{2,i}V_{1,j} - V_{2,j}V_{1,i}}\over{(V_{1,j} - V_{1,i}) - (V_{2,j} - V_{2,i})}},
\end{equation}

\noindent{where V$_1$ and V$_2$ are the measured velocities of the primary and secondary respectively.}

We observed 2 SBs in our northern \emph{PYC} subsample and measured the individual velocities of both components using two-dimensional (2D) cross-correlation techniques.  Target spectra are correlated against pairs of template spectra which are combined at specific flux ratios and over a range of velocity separations.  The primary and secondary velocities appear as peaks in the calculated correlation surface.  RV and $v$sin$i$ measurements presented here were performed using software written in IDL (C. Bender, private communication).  

\begin{figure*}[!h]
\begin{center}
\includegraphics[width=5.3truein]{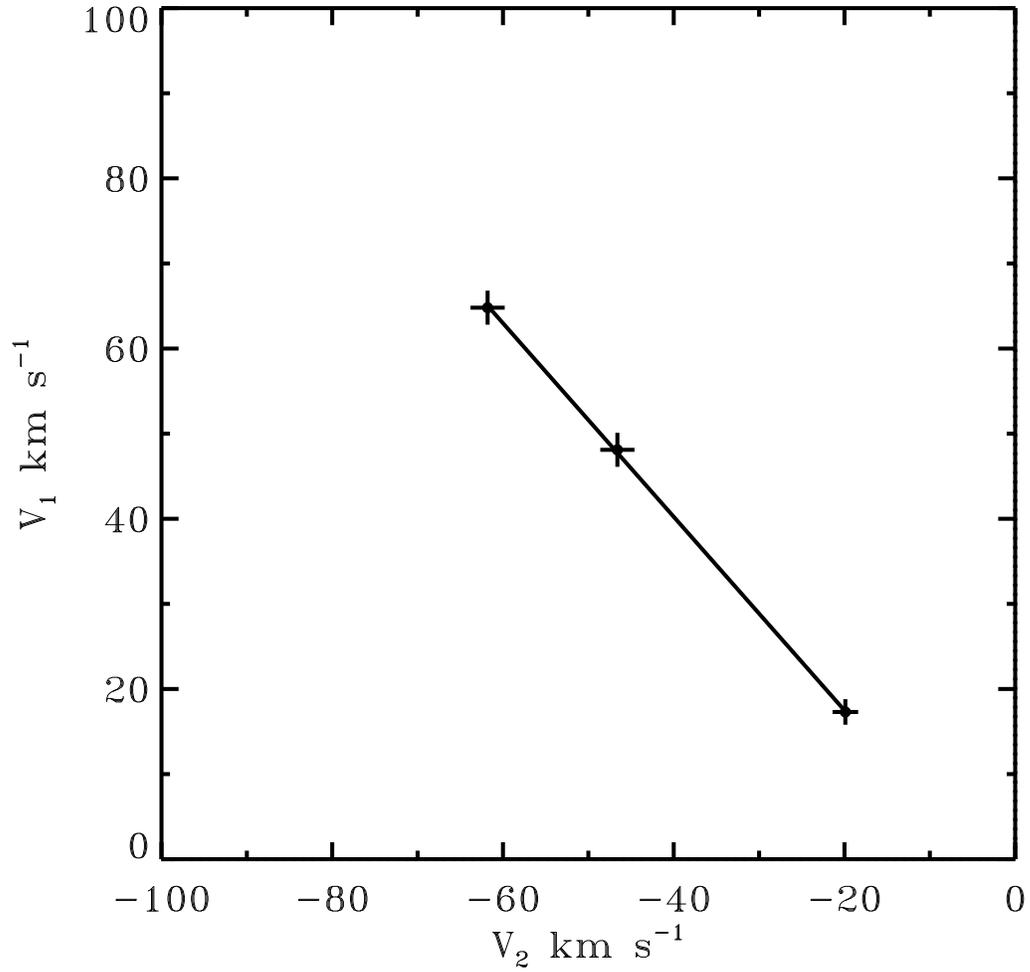}
\caption[HIP 47311AB Component Velocities]{Component velocities of the SB2 HIP 47311AB.  The mass ratio $q$ is equivalent to the negative of the slope of the best fit line to the data (see Equation~1).  The measured $q$ is 1.1$\pm$0.2, indicating the components are equal mass.  The center-of-mass velocity of the system is -2.5$\pm$1.0 km s$^{-1}$.  This velocity is consistent with that predicted for $\beta$ Pic group membership.}
\label{SB_PLOT}
\end{center}
\end{figure*}

\begin{figure*}[!h]
\begin{center}
\includegraphics[width=5.3truein]{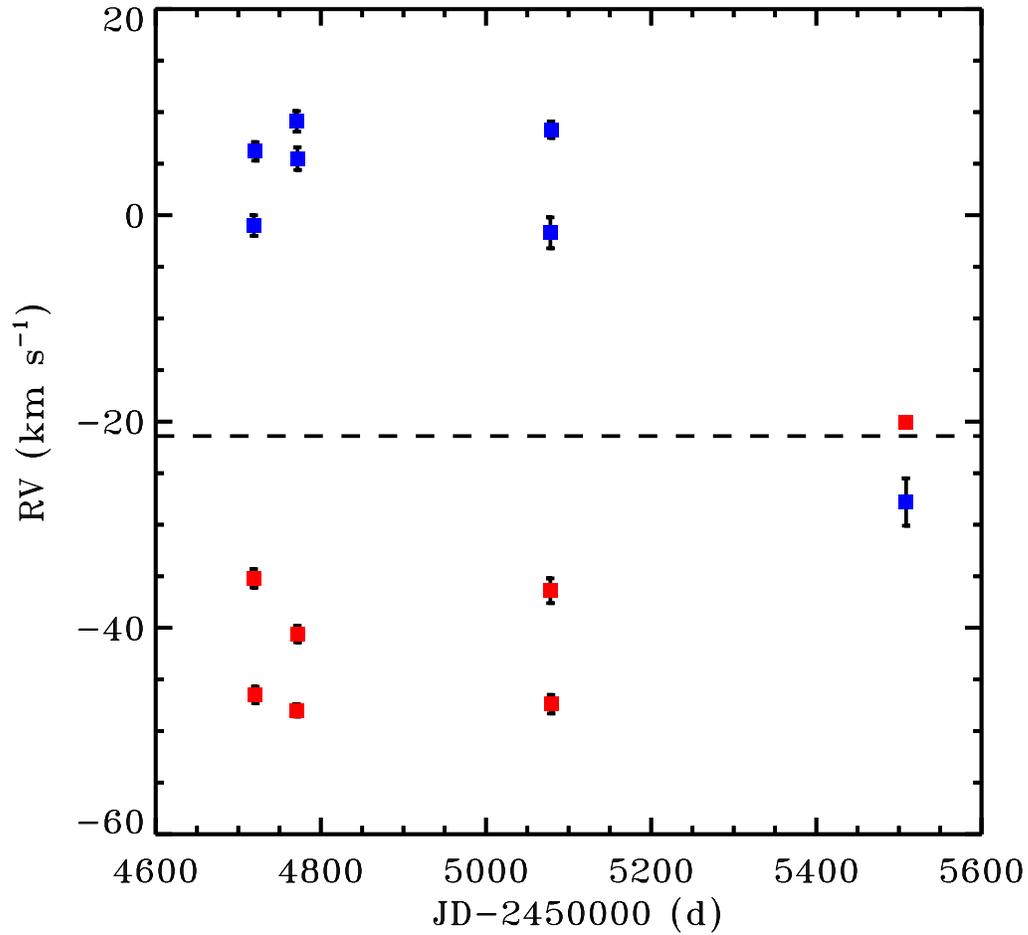}
\caption[PYC 22187AB RV VS. Observation Date]{Component RV as a function of observation date for the PYC 22187+3321AB system.  The primary is shown in red and the secondary in blue. We estimate a mass ratio of 0.84$\pm$0.20, which is consistent with the measured SpTy of the components.  The center-of-mass velocity of the system (black, dashed line) is -21.4$\pm$2.0 km s$^{-1}$.  This velocity is inconsistent with $\beta$ Pic membership and the system is rejected as a likely member.\\ \\(A color version of this figure is available in the online journal.)}
\label{SB_PLOT_2}
\end{center}
\end{figure*}

We acquired 3 spectra of HIP 47311AB from 01/2010 to 05/2011 and 7 spectra of PYC J22187 +3321AB (PYC 22187AB) from 09/2008 to 11/2010.  When treated as single stars during cross-correlation, their RVs vary by as much as 30 km s$^{-1}$ and 9 km s$^{-1}$ respectively over time scales as small as one day.  When treated as double-lined systems in a 2D cross correlation, both stars are consistent with being pairs of M dwarfs.  We identify HIP 47133AB as an M2/M2 system and PYC 22187AB as an M2/M4 system.  We compile the component RV data and observation dates in Table \ref{SB_DATA}.  The component velocities of HIP 47311AB and a linear best fit are shown in Figure~\ref{SB_PLOT}.  In Figure~\ref{SB_PLOT_2} we plot RV data for the components of PYC 22187AB as a function of observation date.  The primary (M2) is shown in red and secondary (M4) in blue.      

 We use equations 1 and 2 to calculate the mass ratio and center-of-mass velocities for these systems.  For HIP 47311AB we calculate $q$ = 1.1$\pm$0.2 and $\gamma$ = -2.5$\pm$1.0 km s$^{-1}$.  This velocity is consistent with $\beta$ Pic membership.  When the RV is combined with the \emph{Hipparcos} parallax (VanLeeuwen~2007), the (\emph{UVW}) velocities of the system match the $\beta$ Pic moving group (see Table~\ref{6D_DATA}).  The measured RVs of the primary and secondary are indicative of a circular orbit.  Furthermore, the very large velocity variation observed over a $\sim$24h period for this system is early evidence of a very tight binary with a short orbital period (see Table~\ref{SB_DATA}). Thus, we cannot exclude the possibility that the observed strong activity is a result of tidal spin-up in the system and it may not be young.  However, we conclude that the pair are likely new members of $\beta$ Pic based on consistent galactic kinematics.  Continued RV monitoring will yield enough data to precisely determine the orbital parameters and provide component mass estimates.  These measurements are critical for direct comparison to evolutionary models and final membership determination.  

We calculate $q$ = 0.84$\pm$0.2 and $\gamma$ = -21.4$\pm$2.0 km s$^{-1}$ for PYC 22187AB.  We show the center-of-mass velocity of the system as a dashed black line in Figure~\ref{SB_PLOT_2}.  This RV rules it out as a likely member of the $\beta$ Pic group.  LG11 estimate a photometric distance of 21.9$\pm$6.6 pc.  Although this estimate is based on unresolved photometry, the true distance to the pair is likely much less than the kinematic estimate of $\sim$69 pc from SLS12.  Orbital parameters from further RV monitoring will provide more information on this likely young, nearby binary.

\section{Discussion}

\subsection{Proposed Member Distances}

It is important to reiterate that our kinematic analyses used the $d_{kin}$ for all candidates.  This distance is an estimate assuming group membership.  We have verified that these distances are consistent with group cluster sequences in color-magnitude diagrams (see SLS12) and consistent with models at the ages of the groups.  For most proposed new members, photometric or spectroscopic distance estimates are available in the literature, but have varying levels of uncertainty (see \S\ref{NORTH_LNM}).  Since the (\emph{UVWXYZ}) calculations are directly dependent on stellar distance, trigonometric parallax measurements from ongoing programs\footnote{Follow-up parallax observations are in progress at the MDM observatory (S. L\'epine, private communication) and US Naval Observatory (H. Harris, private communication).} will determine whether or not \emph{LNM} kinematics are truly consistent with group membership.

\subsection{Proposed Member Ages}

In SLS12 we defined two types of candidate based on their (\emph{V-K$_{s}$}) colors.  We designated candidates with (\emph{V-K$_{s}$})$\le$5.0 candidates with reliable youth (\emph{CWRYs}), and candidates with (\emph{V-K$_{s}$})$>$5.0, candidates with ambiguous youth (\emph{CWAYs}).  These designation were designed to reflect the observed and modeled steep transition in M dwarf activity (H$\alpha$) lifetimes at approximately M4 SpTy (see W11, RM12).  Stars earlier than M4 are active only when young and stars at later types have increasing activity lifetimes, up to $\sim$8 Gyr for the latest M's.  Thus, the activity used as the primary indicator of youth in the SLS12 list has varying levels of reliability depending on candidate spectral type.  Here we discuss the implications of this and other age diagnostics on our proposed members.

Our activity based youth selection is modeled after the methods of Shkolnik et al.~(2009, 2011).   They estimate that X-ray and UV activity consistent with other, young late-type stars places an upper limit of $\sim$300 Myr on the age of an M dwarf. This limit is more reliable for the stars that qualify as \emph{CWRYs}; all but four of the proposed new members.   We can place an age constraint of $<$100 Myr on two of the \emph{CWAY} new members,  PYC J05019+0108 and PYC J21376+0137, based on weak Na 8200~\AA~doublets (see S12).  

Other searches for new \emph{NYMG} members have used the EW of the Li feature at 6708~\AA~as a requirement for membership.  For example, Shkolnik et al.~(2011, 2012) used Li EW as cut in their search for new, low-mass members of moving groups.  They found fewer new, M dwarf members than anticipated and proposed that Li depletion may be too strict a requirement for membership.  Li depletion, like activity, has been shown both empirically and theoretically to be unreliable when estimating low-mass star ages (see Yee and Jensen~2010, Baraffe and Chabrier~2010).  Furthermore, for the case of the older AB Doradus moving group, most M dwarfs will surely have depleted their lithium (see Mentuch et al.~2008)  

Thus, the stars we propose as likely new members of the $\beta$ Pic and AB Dor groups are pre-main sequence, but whether their ages meet the strict criteria imposed by higher mass group members is difficult to determine at this time.  Ongoing spectroscopic follow-up to investigate Li depletion and gravity sensitive spectral features must be combined with existing measurements of photometry, activity, and rotation to build an overarching view of youth on a star by star basis.  The stars we are trying to identify as new moving group members are exactly those stars that elude accurate and reliable age dating. Therefore, the nearby, young M dwarfs presented here are critical to lay the groundwork for new, more reliable methods of age dating low-mass stars.

 \section{Summary}

We present RV follow-up of 27 young, late-type systems in the list of northern $\beta$ Pic and AB Dor moving group candidates from SLS12.  From high resolution near-IR spectra we identify 15 stellar systems having RVs consistent with those predicted for moving group membership.  Kinematic analysis of these stars shows that all are consistent with the (\emph{UVW}) velocity distributions of their respective moving group.  In (\emph{XYZ}) distance space, several of these systems have larger \emph{Z} distances than previously observed for group members.  However, they occupy a region of the sky where members of the $\beta$ Pic and AB Dor groups were not previously known.  We therefore designate all RV match candidates as likely new moving group members; eleven stars in the $\beta$ Pictoris moving group and eleven in the AB Doradus moving group. 

Details on two new spectroscopic binaries are also presented.  One of the binaries, HIP 47133AB, is determined to be a likely new member of the $\beta$ Pic group.  The other, PYC 22187AB, is rejected as a $\beta$ Pic member on the basis of an inconsistent center-of-mass velocity.  These two systems provide the opportunity to measure orbital parameters of nearby, active M dwarfs, yielding component mass estimates, for direct comparison to evolution models.

In total, our search technique and follow-up has thus far proposed 22 likely members of the $\beta$ Pictoris moving group and 22 likely members of the AB Doradus moving group (LS09, SLS10, SLS12, this work).  All of these proposed members are later than SpTy K2, the majority of them being M dwarfs.  Our latest identifications include 12 in binary systems.  If these additions to the proposed membership lists of $\beta$ Pic and AB Dor are determined to be bona-fide members with further follow-up, we will have nearly doubled the under sampled, low-mass populations in both groups.  This work has contributed to the known census of nearby, pre-main-sequence stars and provided excellent targets for disk, exoplanet, and brown dwarf direct imaging studies.  The proposed members also provide well characterized targets for studies of astrophysics in low-mass stars across a range of masses and ages.  We anticipate that continued follow-up of the SLS12 list, and the efforts of others listed in \S\ref{intro}, will provide enough low-mass members to begin concentrated studies of the mass functions and multiplicity fractions of the $\beta$ Pic and AB Dor groups.  Such studies could shed light on their formation histories and provide information about recent local star formation.

\acknowledgments

We thank the anonymous referee for a prompt review that improved the quality of this manuscript. We thank C. Bender for sharing with us his software for the extraction of high-resolution spectra and for their analysis by correlation techniques. J.E.S thanks Tom Herbst for helpful discussions.  We also thank the telescope operators and support astronomers of the NASA-IRTF and \emph{Gemini South}.  The work of J.E.S while at Stony Brook, and of M.S., was supported in part by NSF grant AST 09-07745.  The work of S.L. was supported by NSF grants AST 06-07757 and AST 09-08406.  J.E.S and M.S. were visiting astronomers at the NASA Infrared Telescope Facility, which is operated by the University of Hawaii under Cooperative Agreement no.~NNX-08AE38A with the National Aeronautics and Space Administration, Science Mission Directorate, Planetary Astronomy Program.  Some of these results are based on observations obtained at the Gemini Observatory, which is operated by the Association of Universities for Research in Astronomy, Inc., under a cooperative agreement with the NSF on behalf of the Gemini partnership: the National Science Foundation (United States), the Science and Technology Facilities Council (United Kingdom), the National Research Council (Canada), CONICYT (Chile), the Australian Research Council (Australia), Minist\'{e}rio da Ci\^{e}ncia, Tecnologia e Inova\c{c}\~{a}o (Brazil) and Ministerio de Ciencia, Tecnolog\'{i}a e Innovaci\'{o}n Productiva (Argentina).  This publication makes use of data products from the Two Micron All Sky Survey, which is a joint project of the University of Massachusetts and the Infrared Processing and Analysis Center/California Institute of Technology, funded by the National Aeronautics and Space Administration and the National Science Foundation.  This research has made use of the SIMBAD database, Aladin, and Vizier, operated at CDS, Strasbourg, France.

\clearpage

\begin{sidewaystable*}[!htb]
\scriptsize
\caption[Northern \emph{PYC} Data]{Northern \emph{PYC} Data}
\begin{tabular}{llrrrrrrrrrcr}
\hline
\hline
\emph{PYC} ID			& Alt. ID		         &$\alpha(ICRS)$	&$\ $ $\delta(ICRS)$  & $V\ \ $  &$K_{s}\  $& d$_{kin}$\ \ \ \ & d$_{mod}$\ \ \ \           &RV$_{p}$\ \ \ \      & RV$_{m}$\ \ \ \    &$v$sin$i$\ \ \ \     & SpTy$^a$     & RV\ \ \\ 
                                               &			         & (2000.0)  		&(2000.0)                     &($mag$)       &($mag$)   &($pc$)\ \ \ \          &($pc$)\ \ \ \     &($km~s^{-1}$)       &($km~s^{-1}$)       &($km~s^{-1}$)      &                & OK?  \\
\hline
\multicolumn{13}{c}{$\beta$ Pic Moving Group \emph{PYCs}}\\
\hline
J00325+0729AB$^b$  & LP 525-39AB      &8.145000                 &7.490833                   &12.8        &7.5      &41.1$\pm$4.4 &34.1$\pm$17.0       &1.7$\pm$1.0      &-2.9$\pm$1.1    &20$\pm$2         &M4          &YES\\
J05019+0108  &                                             &75.485625               &1.145250                  &13.2         &7.7      &27.0$\pm$3.2  &24.0$\pm$12.0    &17.1$\pm$1.9   &18.7$\pm$0.5   &8$\pm$2            &M4          &YES\\
J07295+3556AB  &                                             &112.379500             &35.933389                &12.1        &7.8      &42.2$\pm$4.0  &52.3$\pm$26.1      &6.5$\pm$2.0      &10.4$\pm$0.9    &15$\pm$2        &M4         &YES\\
J08227+0757  &                                             &125.697875             &7.954778                  &14.6        &9.2      &62.4$\pm$9.8  &$\cdots$      &12.3$\pm$1.7    &21.3$\pm$0.7    &40$\pm$5        &M4         &NO\\
J09362+3731AB$^c$  &HIP 47133AB               &144.066583             &37.529500                &11.1        &7.2      &33.0$\pm$2.9  &53.0$\pm$26.5     &1.0$\pm$1.6       &-2.5$\pm$1.0     &6$\pm$2          &M2/M2$^c$  &YES\\
J10019+6651$^d$  &                                             &150.499792             &66.857694                &12.8        &8.2      &38.7$\pm$4.0  &43.4$\pm$21.7      &-7.1$\pm$1.5       &-6.2$\pm$0.7    &12$\pm$2        &M3         &YES\\
J10359+2853  &                                             &158.988500             &28.892111                &13.5        &8.4      &37.8$\pm$3.9 & $\cdots$    &-0.4$\pm$1.3     &8.7$\pm$0.4      &4$\pm$2           &M4          &NO\\
J10596+2526E  &BD+26 2161B                 &164.911167             &25.437139                &9.2          &6.0      &22.6$\pm$2.0 &  25.1$\pm$12.6   &-1.2$\pm$1.2     &-3.5$\pm$0.3     &6$\pm$2          &K5           &YES\\
J21185+3014  &TYC 2703 706 1               &319.640667             &30.242944                &11.9        &7.8      &50.5$\pm$8.0  & $\cdots$      &-15.1$\pm$0.9      &-22.5$\pm$0.5  &10$\pm$2          &M2        &NO\\
J21376+0137  & EUVE J2137+01.6           &324.417458             &1.620500                  &13.6       &7.9       &39.2$\pm$4.0 &  24.5$\pm$12.3    &-10.7$\pm$1.2    &-15.1$\pm$5.0  &45$\pm$5        &M5         &YES\\
J22187+3321AB$^c$  &                                &334.677750             &33.353806                &13.4        &8.4       &69.0$\pm$12.0  & $\cdots$  &-12.0$\pm$0.8     &-21.7$\pm$1    &8$\pm$2           &M2/M4$^c$   &NO\\
J22571+3639  &                                             &344.297125             &36.662556                &12.5        &8.6       &68.7$\pm$11.4   &  $\cdots$  &-10.0$\pm$0.9       &-20$\pm$1.2      &20$\pm$5         &M3         &NO\\
\hline
\multicolumn{13}{c}{AB Doradus Moving Group \emph{PYCs}}\\
\hline
J00489+4435AB$^b$  & LP 193-584AB    &12.242667                  &44.585833                 &13.9  &8.2         &32.6$\pm$2.2  & 27.8$\pm$5.6    &-14.2$\pm$1.3   &-15.7$\pm$0.9  &15$\pm$2  &M4  &YES\\
J04065+5628     &TYC 3726 1068 1            &61.645458                   &56.473278                &11.6  &8.4          &51.0$\pm$4.0 & 58.5$\pm$11.7     &-9.6$\pm$1.3       &-11.9$\pm$0.5  &6$\pm$2  &K7  &YES\\
J05599+5834$^{g}$     &GJ 3372B                          &89.982042                   &58.571000                &13.8  &8.2          &24.3$\pm$1.3  & $\cdots$   &-8.9$\pm$1.3       &-1.6$\pm$0.6     &6$\pm$2  &M4  &NO\\
J07080+5816AB$^f$  &3792 2269 1AB                       &107.007792                &58.271361                &11.3  &8.1          &39.9$\pm$2.6  & $\cdots$     &-9.1$\pm$1.3    &38.3$\pm$0.5    &6$\pm$2  &K5  &NO\\
J07102+6218     &TYC 4115 1366 1            &107.553912                &62.313556                &11.8    &8.5        &63.2$\pm$5.9 & $\cdots$     &-11.1$\pm$1.4    &4.1$\pm$0.6      &2$\pm$2  &K5  &NO\\
J08414+6425     & NLTT 19987                    &130.374583                 &64.417528              &12.4  &8.4           &39.9$\pm$2.8 & 42.1$\pm$8.4    &-13.6$\pm$1.4    &-9.6$\pm$0.7    &2$\pm$2  &M2  &YES\\
J10019+6651$^d$  &                                             &150.499792             &66.857694                &12.8        &8.2      &51.3$\pm$4.5  & $\cdots$      &-16.7$\pm$1.4       &-6.2$\pm$0.7    &12$\pm$2        &M3         &NO\\
J10043+5023$^h$     &G 196-3A                          &151.089542                 &50.387111               &11.9  &7.2          &27.7$\pm$1.5  & $\cdots$     &-10.6$\pm$1.5    &-2.8$\pm$0.9    &20$\pm$5  &M4  &NO\\
J10470+4900     &TYC 3446 198 1             &161.763541                &49.001500              &11.0    &7.8         &59.0$\pm$5.3 & $\cdots$    &-12.2$\pm$1.5    &-3.1$\pm$1.4    &0$\pm$2  &K5  &NO\\
J11208+5410      &TYC 3825 716 1             &170.210250                   &54.169111              &12.3  &8.6          &57.9$\pm$5.5  &  56.4$\pm$11.3    &-15.8$\pm$1.5    &-11.1$\pm$0.6  &6$\pm$2  &K7  &YES\\
J12376+3450      &TYC 2533 1877 1          &189.420167                 &34.848861                &12.1  &8.5           &69.4$\pm$7.4  &   57.7$\pm$11.5    &-14.7$\pm$1.6     &-9.7$\pm$0.6    &8$\pm$2  &K7  &YES\\
J13272+4558     &TYC 3460 416 1            &201.800792                 &45.974111                  &11.3  &8.0         &46.0$\pm$4.3  & 49.5$\pm$9.9    &-21.0$\pm$1.6     &-17.6$\pm$2.0     &10$\pm$2  &K7  &YES\\
J13351+5039S  &                                       &203.789417                 &50.654667                  &12.5  &8.4         &52.8$\pm$5.8  & 39.0$\pm$7.8$^e$    &-22.3$\pm$1.5    &-18.9$\pm$1.6  &20$\pm$5  &M3  &YES\\
J13351+5039N  &                                      &203.789413                 &50.654659                    & $\cdots$           &9.4       &52.8$\pm$5.8  & $\cdots$      &-22.3$\pm$1.5    &-17.4$\pm$1.0    &15$\pm$2 &M4  &YES\\
J16458+0343AB  &                                      &251.460917                  &3.717083                   &12.6    &8.4        &50.3$\pm$4.9  & 54.7$\pm$10.9    &-20.6$\pm$1.4    &-15.5$\pm$0.7  &10$\pm$2  &M3  &YES\\
\hline
\multicolumn{13}{l}{\tiny{$^a$determined via cross-correlation with template standards; uncertainty is $\pm$1 sub-class}}\\
\multicolumn{13}{l}{\tiny{$^b$spectroscopic observations were of both components simultaneously}}\\
\multicolumn{13}{l}{\tiny{$^c$spectroscopic binaries, see discussion in Section \ref{SB2}}}\\
\multicolumn{13}{l}{\tiny{$^d$candidate of both $\beta$ Pic and AB Dor}}\\
\multicolumn{13}{l}{\tiny{$^e$based on $K_s$ mag. upper limit in \emph{2MASS}}}\\
\multicolumn{13}{l}{\tiny{$^f$$\sim$3$^{\prime\prime}$ visual binary discovered in CSHELL acquisition images}}\\
\multicolumn{13}{l}{\tiny{$^g$identified as a possible member of the Castor moving group by Shkolnik et al.~(2012)}}\\
\multicolumn{13}{l}{\tiny{$^h$(\emph{UVW}) kinematics do not match any of the moving groups investigated by Shkolnik et al.~(2012)}}\\
\label{PYC_NORTH}
\end{tabular}
\end{sidewaystable*}


\clearpage

\begin{sidewaystable*}[!h]
\caption[Likely New Member Galactic Kinematics]{Likely New Member Galactic Kinematics}
\begin{tabular}{lrrrrrrrr}
\hline
\emph{PYC} ID			  &$\mu$$_{\alpha}$\ \ \ \ \ \ \	            & $\mu$$_{\delta}$\ \ \ \ \ \ \                      & $U\ \ \ \ \ $               &$V\ \ \ \ \ $                 & $W\ \ \ \ \ $                   & $X\ \ \ \ $                &$Y\ \ \ \ $                   & $Z\ \ \ \ $    \\
                                                 &($mas~yr^{-1}$)                           &($mas~yr^{-1}$)                                    &($km~s^{-1}$)         &($km~s^{-1}$)          &($km~s^{-1}$)             &($pc$)\ \ \ \              &($pc$)\ \ \ \                &($pc$)\ \ \ \      \\
\hline
\multicolumn{9}{c}{$\beta$ Pic Moving Group \emph{LNMs}}\\
\hline
J00325+0729AB    &92.0$\pm$8.0  &-55.0$\pm$8.0   &-9.8$\pm$1.7  &-18.0$\pm$1.7   &-5.0$\pm$2.3   &-9.9$\pm$0.3   &21.4$\pm$2.1  &-33.7$\pm$3.7 \\
J05019+0108   &23.0$\pm$8.0  &-92.0$\pm$8.0 &-11.2$\pm$1.2  &-16.1$\pm$1.6  &-10.6$\pm$1.5  &-23.5$\pm$2.6   &-7.9$\pm$1.3 &-10.8$\pm$0.6 \\
J07295+3556  &-30.0$\pm$8.0  &-95.0$\pm$8.0  &-13.2$\pm$2.0  &-16.9$\pm$1.8   &-6.6$\pm$1.5  &-38.9$\pm$2.1   &-2.0$\pm$1.5   &16.3$\pm$0.8 \\
J09362+3731AB  &-99.4$\pm$2.5  &-89.5$\pm$1.4   &-8.7$\pm$1.7  &-15.2$\pm$1.1  &-12.6$\pm$0.7  &-22.4$\pm$0.8   &-2.3$\pm$1.0   &25.0$\pm$1.2 \\
J10019+6651  &-74.0$\pm$8.0  &-78.0$\pm$8.0   &-9.3$\pm$2.2  &-17.4$\pm$1.5   &-6.3$\pm$1.3  &-23.0$\pm$1.6   &16.7$\pm$2.5   &26.2$\pm$1.6 \\
J10596+2526E  &-177.5$\pm$1.7  &-81.5$\pm$1.6  &-11.9$\pm$1.7  &-13.1$\pm$0.5  &-11.8$\pm$0.8   &-8.3$\pm$0.2   &-4.9$\pm$0.8   &20.5$\pm$1.3 \\
 J21376+0137  &83.0$\pm$8.0  &-54.0$\pm$8.0  &-13.3$\pm$2.7  &-18.7$\pm$3.7   &-6.3$\pm$3.5   &17.6$\pm$1.2   &26.7$\pm$1.4  &-22.6$\pm$2.3 \\
\hline
\multicolumn{9}{c}{AB Dor Moving Group \emph{LNMs}}\\
\hline
J00489+4435AB   &116.0$\pm$8.0  &-136.0$\pm$8.0   &-3.4$\pm$1.5  &-27.7$\pm$1.2  &-15.2$\pm$1.9  &-16.6$\pm$0.5   &26.1$\pm$1.0  &-10.2$\pm$1.0 \\
J04065+5628   &38.0$\pm$8.0  &-116.0$\pm$8.0   &-4.4$\pm$1.8  &-27.6$\pm$2.2  &-15.3$\pm$2.3  &-43.0$\pm$1.6   &27.3$\pm$2.1    &2.9$\pm$1.3 \\
J08414+6425  &-70.0$\pm$8.0  &-130.0$\pm$8.0  &-10.1$\pm$1.3  &-25.1$\pm$2.3  &-11.8$\pm$1.4  &-28.3$\pm$1.1   &15.5$\pm$1.7   &23.5$\pm$1.1 \\
J11208+5410  &-81.3$\pm$6.3   &-53.5$\pm$5.8  &-11.4$\pm$1.1  &-24.1$\pm$2.8  &-11.2$\pm$1.8  &-25.9$\pm$1.8   &16.0$\pm$2.8   &49.2$\pm$2.3 \\
J12376+3450  &-71.0$\pm$8.0   &-43.0$\pm$8.0  &-11.6$\pm$1.6  &-25.0$\pm$3.6   &-9.0$\pm$2.6   &-8.0$\pm$1.2    &6.0$\pm$3.0   &68.7$\pm$4.4 \\
J13272+4558  &-90.3$\pm$2.3  &-54.2$\pm$2.1   &-6.9$\pm$1.7  &-26.2$\pm$2.1  &-10.1$\pm$1.4   &-4.0$\pm$1.1   &15.3$\pm$1.9   &43.2$\pm$1.9 \\ 
J13351+5039SN  &-81.0$\pm$8.0  &-30.0$\pm$8.0   &-8.9$\pm$1.7  &-25.1$\pm$2.8  &-10.8$\pm$2.3   &-6.3$\pm$1.7   &21.4$\pm$2.8   &47.9$\pm$2.3 \\
J16458+0343AB  &-32.0$\pm$8.0  &-92.0$\pm$8.0   &-2.5$\pm$1.6  &-25.5$\pm$2.7  &-11.2$\pm$1.7   &40.9$\pm$3.8   &15.7$\pm$2.0   &24.8$\pm$1.0 \\
\hline
\hline
\label{6D_DATA}
\end{tabular}
\end{sidewaystable*}

\clearpage

\begin{table*}[!h]
\begin{center}
\caption[SB2 RV Data]{SB2 RV Data}
\begin{tabular}{lrr}
\hline
\hline
JD-2450000	 &$V_1$\ \ \ \	  		&$V_2$\ \ \ \ \\
			& $km~s^{-1}$  		& $km~s^{-1}$ 	\\
\hline
\multicolumn{3}{c}{HIP 47311AB}\\
\hline
5211.993056  &48.1$\pm$2.0  &-46.6$\pm$2.0\\
5212.972222  &17.3$\pm$1.5  &-19.9$\pm$1.5\\
5685.736110  &64.8$\pm$2.0  &-61.8$\pm$2.0\\
\hline
\multicolumn{3}{c}{PYC J22187+3321AB}\\
\hline
4718.868056  &-35.2$\pm$0.9  &-1.0$\pm$1.0\\
4720.770833  &-46.5$\pm$0.8  &6.2$\pm$0.9\\
4770.813194  &-48$\pm$0.6  &9.1$\pm$1.0\\
4771.763889  &-40.6$\pm$0.8  &5.5$\pm$1.1\\
5077.914583  &-36.4$\pm$1.2  &-1.7$\pm$1.5\\
5078.824306  &-47.4$\pm$0.9  &8.3$\pm$0.8  \\
5508.736806  &-20.1$\pm$0.4  &-27.8$\pm$2.3\\
\hline
\label{SB_DATA}
\end{tabular}
\end{center}
\end{table*}

\clearpage 

\clearpage

\clearpage

\clearpage

\clearpage

\clearpage

\end{document}